\documentclass[runningheads]{llncs}
\usepackage[T1]{fontenc}
\usepackage{graphicx}
\usepackage{listings}
\usepackage{amsmath}
\usepackage{amssymb}
\usepackage{graphicx}
\usepackage{bm}
\usepackage{color}
\usepackage{authblk}
\usepackage{booktabs}
\usepackage{multirow}
\usepackage{float}
\usepackage{tikz}
\usepackage{ctable}
\usepackage[colorlinks, citecolor=green]{hyperref}
\newcommand\tstrut{\rule{0pt}{2.4ex}}

\usepackage{color, colortbl}
\definecolor{Gray}{gray}{0.9}
\usetikzlibrary{arrows.meta,arrows}
\usepackage{diagbox}
\usepackage{breakcites}
\usepackage{orcidlink}
\usepackage{setspace}

\begin{document}
\title{Whole Heart 3D+T Representation Learning Through Sparse 2D Cardiac MR Images}
\titlerunning{Early accepted by MICCAI 2024}

%

\author{
Yundi Zhang\inst{1,2}\orcidlink{0009-0008-7725-6369} \and
Chen Chen\inst{3,5,6}\orcidlink{0000-0002-3525-9755},
Suprosanna Shit\inst{4}\orcidlink{0000-0003-4435-7207} \and 
Sophie Starck\inst{1,2} \and
Daniel~Rueckert\inst{1,2,5}\orcidlink{0000-0002-5683-5889} \and
Jiazhen Pan\inst{1,2}\orcidlink{0000-0002-6305-8117}
}

\authorrunning{Y. Zhang et al.}
%
\institute{School of Computation, Information and Technology, Technical University of Munich, Germany
\and School of Medicine, Klinikum Rechts der Isar, Technical University of Munich, Germany
\and Department of Computer Science, University of Sheffield, UK
\and Department of Quantitative Biomedicine, University of Zurich, Switzerland,
\and Department of Computing, Imperial College London, UK
\and Department of Engineering Science, University of Oxford, UK
\\
\email{\{yundi.zhang,jiazhen.pan\}@tum.de}
}

\maketitle              
\begin{abstract}
Cardiac Magnetic Resonance (CMR) imaging serves as the gold-standard for evaluating cardiac morphology and function. Typically, a multi-view CMR stack, covering short-axis (SA) and 2/3/4-chamber long-axis (LA) views, is acquired for a thorough cardiac assessment. However, efficiently streamlining the complex, high-dimensional 3D+T CMR data and distilling compact, coherent representation remains a challenge. In this work, we introduce a whole-heart self-supervised learning framework that utilizes masked imaging modeling to automatically uncover the correlations between spatial and temporal patches throughout the cardiac stacks. This process facilitates the generation of meaningful and well-clustered heart representations without relying on the traditionally required, and often costly, labeled data. The learned heart representation can be directly used for various downstream tasks. Furthermore, our method demonstrates remarkable robustness, ensuring consistent representations even when certain CMR planes are missing/flawed. We train our model on 14,000 unlabeled CMR data from UK BioBank and evaluate it on 1,000 annotated data. The proposed method demonstrates superior performance to baselines in tasks that demand comprehensive 3D+T cardiac information, e.g. cardiac phenotype (ejection fraction and ventricle volume) prediction and multi-plane/multi-frame CMR segmentation, highlighting its effectiveness in extracting comprehensive cardiac features that are both anatomically and pathologically relevant.

\end{abstract}
\section{Introduction}

Cardiac Magnetic Resonance (CMR) imaging plays an essential role in the evidence-based diagnosis of cardiovascular diseases, establishing itself as the gold-standard~\cite{von2017representation} for cardiac morphology and function assessment by offering detailed 3D+T heart visualization. However, high-resolution 3D+T CMR acquisition is not practical in clinics due to the requirement for long breath-holds and the reduced CMR contrast for further cardiac evaluation. Therefore, multi-view 2D+T CMR imaging including a stack of short-axis (SA) planes and 2/3/4 long-axis (LA) planes is preferable in clinical practice. Nonetheless, how to efficiently process these complex, high-dimensional CMR sequences and seamlessly integrate them into a coherent and unified 3D+T representation for a thorough cardiac assessment lacks a simple solution.


In the past decade, a large multitude of CMR representation learning methods have been proposed for cardiac function analysis~\cite{xue2017direct,biffi2018learning,wang2021joint,qiu2023multimodal,ecg2}. They extract relevant features from CMR images, forming a compressed latent representation space essential for customized tasks. However, four major challenges remain in the field of cardiac representation learning. First, in most studies, the representation learning of cardiac morphology and function relies on curated annotated data. Yet, a vast majority of the CMR images are unlabeled. Only unsupervised/self-supervised (SSL) methods can leverage these large-scale CMR images and construct a comprehensive, meaningful representation. Second, most cardiac representation learning works focus on a specific downstream task, such as super-resolution and different modality mapping. Cardiac representation learning that can be generalized to multiple downstream tasks (e.g. cardiac segmentation and critical phenotype prediction) is still not broadly studied in the community. Third, existing studies demand a rigorous amount of inputs and lack adaptability and robustness to handle incomplete inputs, particularly when certain CMR planes are either not acquired or defective. Lastly, to the best of our knowledge, none of the previous studies efficiently incorporates all available spatial and temporal information from CMR. They either operate exclusively on a limited set of cardiac planes or neglect the utilization of temporal information.

In this paper, we introduce a cardiac representation learning method that is \textbf{self-supervised, scalable to vast unlabeled datasets, adaptable to diverse downstream tasks, flexible in handling varying amounts of input CMR planes, and capable of integrating comprehensive spatiotemporal information from sparse CMR inputs}. Our key contributions are:
\begin{enumerate}
    \item We propose a 3D+T representation learning method for the whole heart, which is learned using multi-view (SA and LA) planes together with temporal information. By exploring correlation across different SA and LA sequences in an SSL manner, our model attains a rich and meaningful cardiac representation that can be adapted to different tasks.
    \item Our approach ensures a consistent cardiac representation of the same subject, even in the absence of a few planes. This attribute is particularly beneficial in practice where some SA/LA planes are not available or of poor quality. This enables the same effectiveness as full CMR scans but with reduced acquisition costs and times.
    \item We train our model with 14,000 unlabeled CMR data from UK-BioBank~\cite{petersen2015uk} and evaluate it on 1,000 annotated data. The visualized meaningful representation, the accurate cardiac phenotype prediction, e.g. ventricle volume and ejection fraction (which can only be achieved by leveraging adequate 3D+T information), and the enhanced end-to-end all-planes CMR segmentation to baselines demonstrate its capability of dealing with various downstream tasks that require a comprehensive understanding of the whole heart. Its manifested versatility paves the way toward a cardiac MR foundation model.
\end{enumerate}
\subsection{Related Work}\label{sec2}

\paragraph{\textbf{Cardiac imaging analysis}} can be conducted in various ways. Cardiac segmentation is the most common way to evaluate cardiac function. The relevant cardiac phenotype and clinic-useful parameters can be extracted based on it~\cite{khened2019fully,bai2020population,chen2020deep,campello2021multi}. However, whole-heart segmentation utilizing multi-view CMR sequences has not been widely studied. \cite{chen2019learning} used multi-view 2D CMR data to learn the correlation between different views and ameliorate segmentation performance but no temporal redundancy was exploited. \cite{qin2018joint} leveraged the temporal redundancy using RNNs and performed the 2D+T segmentation. \cite{stolt2023nisf} reconstructed high-resolution 3D SA segmentation using a neural implicit function but lacked LA knowledge integration. Moreover, critical cardiac phenotype/values can also be extracted directly from CMR without segmentation~\cite{zhen2016multi,LV_pred,xue2017direct}, which side-step the non-trivial manual cardiac imaging annotation. Cardiac motion/mesh tracking~\cite{wang2011cardiac,pan2021efficient,meng2022mulvimotion,pan2024} is another important approach to analyse cardiac morphology. Notably, both phenotype estimation (e.g. ejection fraction) and 3D cardiac motion require ample spatial/temporal information. Utilizing extensive 3D+T information from different views is thus advantageous for these analyses.

\paragraph{\textbf{Cardiac representation learning}} simplifies the high-dimensional and complex CMR data into more manageable forms, enabling models to focus on the most relevant features of the heart. It can be applied to solve/facilitate different pre-defined tasks, e.g. cardiac segmentation~\cite{Factorised,biffi2018learning,sun2022attention}, image reconstruction~\cite{schlemper2018cardiac,pan2023global}, super-resolution~\cite{wang2021joint} or cardiac indices prediction~\cite{xue2017direct}. Notably, these methods predominantly relying on supervised learning face scalability issues in large unannotated datasets. Furthermore, cardiac representation can also bridge information from a different modality, e.g. ECG~\cite{turgut2023unlocking}, genomics~\cite{ecg2}, and radiology reports~\cite{qiu2023multimodal}, enhancing the breadth of data analysis. However, these approaches do not fully utilize the available 3D+T information from multi-view CMR, potentially missing critical spatial or temporal insights. Moreover, it is also challenging to align and correlate the information efficiently from these different views.

\section{Methods}
\begin{figure}[!t]
  \centering
  \includegraphics[width=\textwidth]{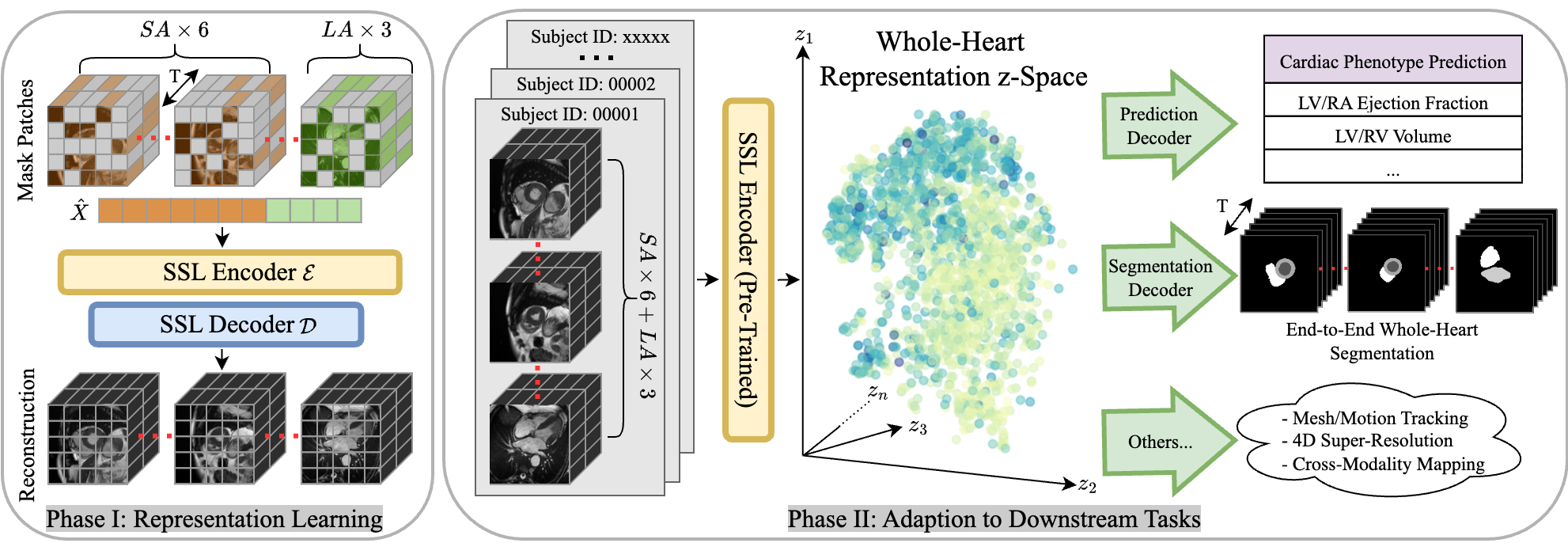}
  \caption{\textbf{Overview of the proposed method.} \underline{Phase I}: Representation learning is achieved through SSL reconstruction of a stack of multi-view masked 2D+T slices (6 SA and 3 LA). \underline{Phase II}: Leveraging the whole-heart latent representation learned from the pre-trained SSL encoder, we utilize various decoders to carry out downstream tasks, e.g. cardiac phenotype prediction and whole-heart segmentation.}
  \label{fig:main}
\end{figure}

We develop our method's backbone based on Masked Autoencoders (MAE)~\cite{he2022mae} due to its alignment with our criteria, including inherent self-supervised learning characteristics, adaptability for various downstream tasks, and robustness in handling missing inputs. To incorporate high-dimensional cardiac information, we rebuild MAE from a simple 2D model to a cardiac representation learning method operating on complex multi-view CMR sequences. The training process has two phases. In phase I, we leverage a huge amount of unlabeled data to learn representations. In phase II, the learned representations can be effortlessly extended to diverse downstream tasks, such as cardiac phenotype prediction and all-planes CMR segmentation. The model's architecture is illustrated in Fig.~\ref{fig:main}.

\paragraph{\textbf{Phase I: Representation learning.}} We propose a simple but effective solution to correlate the "interlaced" multi-view CMR sequences and learn a whole-heart representation from them. We assume that each scan includes a set of sparse 2D+T CMR sequences with $M$ SA planes $\{S_1, ..., S_M\}$ and $N$ LA planes $\{L_1, ..., L_N\}$. Directly stacking all 2D+T sequences into a 4D tensor and conducting a 4D operation is computationally prohibitive. Nevertheless, LA and SA planes are acquired from different views and do not share spatial feature similarity. Thus, instead of forming a 4D tensor we first decompose each plane into $P$ small 2D+T patches, denoted as $s^{m}_{p}$ for the $p$-th patch of $m$-th SA plane and $l^{n}_{p}$ for the $p$-th patch of $n$-th LA plane. Our method takes all these individual patches as an input vector, denoted as $X = [s^{1}_{1}, ..., s^{M}_{P}, l^{1}_{1}, ..., l^{N}_{P}]$, containing in total $(M+N)\times P$ patches. To capitalize on the temporal redundancy inherent in CMR data and further reduce computational demands, we use a larger patch size along the temporal than the spatial dimension.
To enhance localization, each patch is added with a 4D positional embedding~\cite{he2022mae}, indicating the patch's $x\textrm{--}y\textrm{--}t$ spatiotemporal index and the corresponding plane of origin. Then a random mask is applied on the input to mask away $q\%$ of the patches. The shortened input with remaining patches, $\hat{X}$ (refer to Fig~\ref{fig:main}), are then fed into an encoder $\mathcal{E}$ to learn a dense 3D+T cardiac representation. A high maskout ratio $q\%$ is chosen here to force the encoder to find the underlying spatiotemporal correlation across different cardiac planes, rather than easily extrapolate missing pixel intensity. Afterwards, a lightweight decoder $\mathcal{D}$ is applied to predict the masked-out patches and reconstruct $X$. We applied the mean squared error (MSE) for optimization  with a loss function formulated as $\mathcal{L} = \left\Vert X - \mathcal{D}\left(\mathcal{E}\left(\hat{X}\right)\right) \right\Vert_2^2.$

\paragraph{\textbf{Phase II: Adaption to different downstream tasks.}} The 3D+T whole heart representation can be used to solve different downstream tasks. 
We fine-tune our model on a small-scale dataset where corresponding labels are available. In phase II, we forward all available CMR planes to the model without maskout. The encoder remains unchanged and the reconstruction decoder is replaced with a different decoder according to the task we aim to solve. For cardiac phenotype (e.g. ejection fraction and volume) prediction, we use shallow linear layers since we assume the spatiotemporal representation learned in phase I is already comprehensive and can represent cardiac morphology to some extent. For all-planes CMR sequences segmentation, a U-Net-wise decoder with skip-connections~\cite{zhou2023medmae} is adopted. It is noteworthy that, in contrast to conventional cardiac segmentation (refer to section ~\ref{sec2}) which can only conduct specific 2D plane segmentation at one inference time, our model can segment \textbf{all CMR planes/frames} in an end-to-end fashion at a \textbf{single} inference time. Furthermore, since the model can exploit the correlation between different CMR planes, the final segmentation is further enhanced and is expected to perform better than single plane segmentation~\cite{chen2019learning} (more details refer to Table~\ref{tab:seg} and Fig.~\ref{fig:seg}).


\paragraph{\textbf{Robustness when some planes are absent.}} In clinic scenarios, variations in acquisition protocols or artifacts from patient motion can lead to missing/corrupted CMR planes. Our method, however, remains robust against these challenges due to its contrastive learning inherence. Contrastive learning is usually used in representation learning to pull positive (similar) data and push dissimilar data~\cite{byol,chen2021exploring}. Our method employs a form of implicit contrastive learning, aligning positive pairs through diverse mask patterns~\cite{zhang2022mask}. During different training epochs in phase I, our model is forced to generate the same representation/reconstruction for a subject, regardless of different random mask applications. This ensures that even in scenarios where certain CMR planes are missing, our model can still deliver the same representation as complete CMR plane inputs. 

\section{Dataset and Experiments}
\textit{\textbf{Dataset.}}
Our model is trained on the CMR data from UK BioBank~\cite{petersen2015uk}. We use 14,000 subjects to conduct representation learning and finetune the model on different downstream tasks using 1,000 annotated subjects with segmentation masks and cardiac phenotype labels from~\cite{bai2020population}. The tests are performed on 100 subjects. Each subject in the dataset includes 6 SA and 3 LA 2D slices with 50 time frames. All slices are cropped to the cardiac center of size $128 \times 128$.

\noindent \textit{\textbf{Implementation details.}}
We implement 6 encoder layers and 2 decoder layers in SSL with an embedding dimension of 1024. The patch size is $8\times 8\times 25$ with 8 the spatial and 25 the temporal dimension. The mask ratio $q\%$ in SSL is set to $70\%$. For downstream tasks, we set the embedding dimension to 256 for the prediction decoder and 576 for the segmentation decoder. We conduct all experiments on a NVIDIA A6000 GPU. The batch size is set to 1 for representation learning and segmentation, and 4 for cardiac phenotype prediction. The initial learning rate (LR) is set to $10^{-4}$ and we use a cosine scheduler with a weight decay of 0.05. The LR for downstream tasks' fine-tuning is set to $10^{-5}$. 

\noindent \textit{\textbf{Downstream baseline methods.}}
For prediction, we compare our method with ResNet50~\cite{he2016deep} and ViT~\cite{dosovitskiy2020vit}. They both take concatenated 3D+T planes as input and T is treated as channel dimension. For multi-plane segmentation, we compare our method with nnUNet~\cite{isensee2021nnunet} and UNETR~\cite{hatamizadeh2022unetr}. As nnUNet can only take 2D+T planes as inputs, we have to carry out two times training (once for SA and once for LA) and treat different planes separately. We adapt UNETR to UNETR+ so that it can take all 2D+T planes as input at once (therefore also 3D+T), aligning with our setting. Moreover, we compare to our ablated methods which are trained only using LA or SA CMR. This allows us to assess the extent to which performance can be enhanced when additional spatiotemporal information from other planes is incorporated.

\section{Results and Discussion}\label{sec5}
\label{subsec:results}
\paragraph{\textbf{Representation learning evaluation via reconstruction.}} We first evaluate reconstruction using only SA sequences, only LA sequences, and all available views in pre-training. The quantitative results and reconstructed CMR images are shown in Appendix Table~\ref{tab:recon} and Fig.~\ref{fig:recon}. The superior results of all-view reconstruction to that of using only limited spatial planes imply the benefit of involving all-view spatiotemporal information for whole-heart representation learning. Incorporating different views can help the model build spatial knowledge of the heart and therefore enhance the representation/reconstruction.

\begin{figure}[ht]
  \centering
  \includegraphics[width=\textwidth]{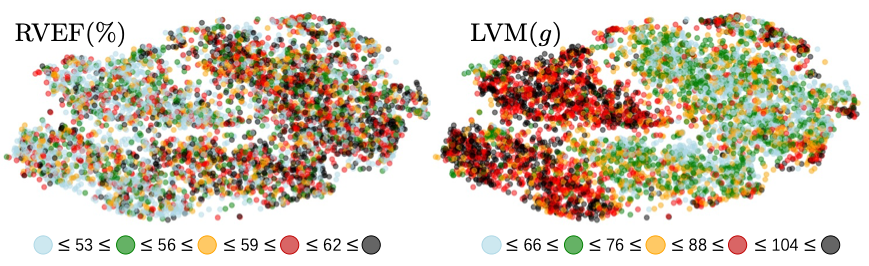}
  \caption{The t-SNE visualization of the latent whole-heart representation obtained through pre-training. Latent embeddings are labeled with RVEF and LVM values, categorized into 5 groups according to the ground truth, and shown in different colors.}
  \label{fig:t-SNE}
\end{figure}

\paragraph{\textbf{Representation learning evaluation via t-SNE visualization.}} We further provide the t-SNE visualization of the learned representations in Fig.~\ref{fig:t-SNE}. We label the latent embeddings after pre-training with right ventricular ejection fraction (RVEF) and left ventricle mass (LVM). The generation of these two phenotypes relies on 3D spatial and temporal information of the heart. Notably, even without using any labels during pre-training, our model can already generate a well-clustered latent space reflective of spatiotemporal differences across the subjects. This visualization serves as strong evidence for the effectiveness of the learned representations.

\begin{table}[ht]
\centering
\setlength{\tabcolsep}{0.4mm}{}
\caption{Comparison of mean absolute errors among mean-guess (estimating every subject's phenotype value with the cohort mean value), ResNet50, ViT, and the proposed approach for predicting LVM, RVEF, RAEF, RVEDV, and LASV. ResNet50 and ViT use concatenated 3D+T planes as input.}
\label{tab:prediction}
\begin{tabular}{l|c|c|c|c|c}
\toprule
Method   & LVM (g)   & RVEF (\%) & RAEF (\%)& RVEDV (mL)& LASV (mL)    \\ \hline
\rowcolor{Gray}
Mean-guess & $17.103_{\pm11.104}$ & $12.14_{\pm6.808}$ & $6.586_{\pm 5.899}$& $31.915_{\pm 22.807}$& $8.918_{\pm 6.113}$   \tstrut\\ 
ResNet50 & $7.150_{\pm6.945}$ & $3.245_{\pm2.354}$ & $6.294_{\pm 4.983}$& $12.577_{\pm 11.003}$& $5.777_{\pm 4.695}$   \tstrut\\ 
\rowcolor{Gray}
ViT      & $8.656_{\pm7.858}$ & $4.292_{\pm3.197}$ & $6.590_{\pm 5.486}$& $14.102_{\pm 12.758}$& $8.944_{\pm 6.219}$   \tstrut\\ 

Ours & $\mathbf{4.332_{\pm4.716}}$ & $\mathbf{2.831_{\pm2.550}}$ & $\mathbf{5.396_{\pm 4.213}}$& $\mathbf{10.713_{\pm 8.980}}$& $\mathbf{3.529_{\pm 3.186}}$   \tstrut\\
\bottomrule
\end{tabular}
\end{table}

\paragraph{\textbf{Cardiac phenotype prediction.}} We assess the efficacy of learned whole-heart representations by predicting critical phenotype values. Table~\ref{tab:prediction} shows the mean absolute error of our predictions compared to ResNet50 and ViT for LVM, RVEF, right atrial ejection fraction (RAEF), right ventricular end-diastolic volume (RVEDV), and left atrial stroke volume (LASV). For accurate prediction of these values, both adequate spatial and temporal cardiac information is essential. The superior performance to the baseline methods underscores its capability to capture high-level spatiotemporal features of the entire heart.



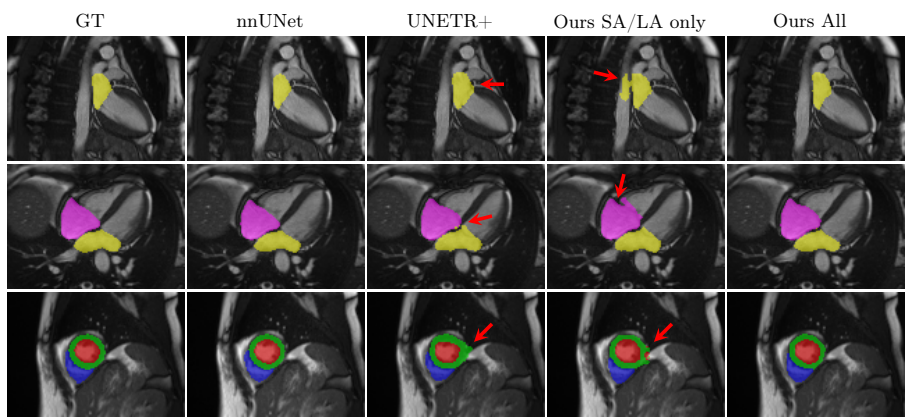
\begin{figure}[ht]
  \centering
  \resizebox{\textwidth}{!}{%
    \begin{tikzpicture}[font=\small]
    \node[anchor=north] at (1.5, 13.2) {GT};
    \node[anchor=north] at (4.5, 13.2) {nnUNet};
    \node[anchor=north] at (7.5, 13.2) {UNETR+};
    \node[anchor=north] at (10.5,13.2) {Ours SA/LA only};
    \node[anchor=north] at (13.5,13.2) {Ours All};
    \node[anchor=south west, scale=1.0] at (0,10.5)
    {\includegraphics[width=2.95cm]{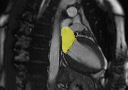}};
    \node[anchor=south west, scale=1.0] at (3,10.5)
    {\includegraphics[width=2.95cm]{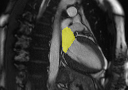}};
    \node[anchor=south west, scale=1.0] at (6,10.5)
    {\includegraphics[width=2.95cm]{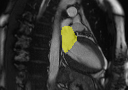}};
    \node[anchor=south west, scale=1.0] at (9,10.5)
    {\includegraphics[width=2.95cm]{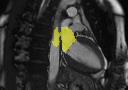}};
    \node[anchor=south west, scale=1.0] at (12,10.5)
    {\includegraphics[width=2.95cm]{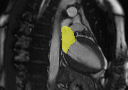}};
    \node[anchor=south west, scale=1.0] at (0,8.36)
    {\includegraphics[width=2.95cm]{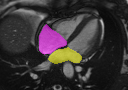}};
    \node[anchor=south west, scale=1.0] at (3,8.36)
    {\includegraphics[width=2.95cm]{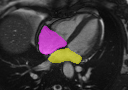}};
    \node[anchor=south west, scale=1.0] at (6,8.36)
    {\includegraphics[width=2.95cm]{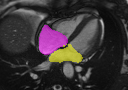}};
    \node[anchor=south west, scale=1.0] at (9,8.36)
    {\includegraphics[width=2.95cm]{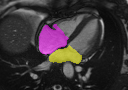}};
    \node[anchor=south west, scale=1.0] at (12,8.36)
    {\includegraphics[width=2.95cm]{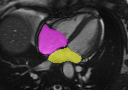}};
    \node[anchor=south west, scale=1.0] at (0,6.23)
    {\includegraphics[width=2.95cm]{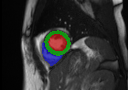}};
    \node[anchor=south west, scale=1.0] at (3,6.23)
    {\includegraphics[width=2.95cm]{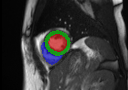}};
    \node[anchor=south west, scale=1.0] at (6,6.23)
    {\includegraphics[width=2.95cm]{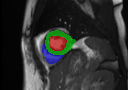}};
    \node[anchor=south west, scale=1.0] at (9,6.23)
    {\includegraphics[width=2.95cm]{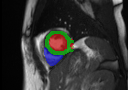}};
    \node[anchor=south west, scale=1.0] at (12,6.23)
    {\includegraphics[width=2.95cm]{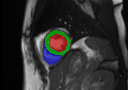}};
    \draw [-stealth,red,line width=0.6mm] (8.2,7.9) -- (7.9,7.6);
    \draw [-stealth,red,line width=0.6mm] (11.2,7.9) -- (10.9,7.6);
    \draw [-stealth,red,line width=0.6mm] (8.2,9.7) -- (7.8,9.6);
    \draw [-stealth,red,line width=0.6mm] (10.4,10.4) -- (10.3,10);
    \draw [-stealth,red,line width=0.6mm] (8.4,11.9) -- (8,11.9);
    \draw [-stealth,red,line width=0.6mm] (9.9,12.1) -- (10.3,12);
    \end{tikzpicture}
    }
  \caption{Qualitative segmentation results among nnUNet, UNETR+, and the proposed methods. UNETR+ and the proposed approach in the last column (Ours All) use all sparse CMR sequences as network input, while nnUNet and the second last column (Ours SA/LA) are trained solely with either SA or LA views.}
  \label{fig:seg}
\end{figure}

\paragraph{\textbf{Multi-view segmentation.}}
The competence of our model to perform end-to-end segmentation across all planes is shown in Table~\ref{tab:seg} and Fig.~\ref{fig:seg}. Our model not only exhibits comparable quantitative dice scores with nnUNet (that leverages exhaustive parameters tuning) but also shows superior performance over UNETR+ in all regions for both SA/LA planes. This benefit is gained from the learned whole-heart representation. Moreover, the superior performance against SA-only and LA-only segmentation also highlights the significance of integrating multi-view CMR information for more accurate segmentation outcomes.

\begin{table}[ht]
\centering
\setlength{\tabcolsep}{2.2mm}{}
\caption{Segmentation dice scores of nnUNet, UNETR+, and proposed methods trained with SA-only, LA-only and all-planes CMR. nnUNet employs a single 2D+T plane as input (therefore 2 times training for SA/LA) while UNETR+ uses all sparse CMR sequences as input, same as ours. The top two results are marked in bold.} 
\label{tab:seg}

\begin{tabular}{l|c|c|c|c|c}
\toprule
Phenotype & nnUNet & UNETR+  & Ours (SA) & Ours (LA) & Ours (All)  \\ \hline
\rowcolor{Gray}
LVBP       &$\mathbf{0.98}_{\pm0.01}$&$0.89_{\pm0.02}$&$0.95_{\pm0.02}$& N.A.               &$\mathbf{0.96}_{\pm0.01}$   \tstrut\\ 
LVMYO      &$\mathbf{0.96}_{\pm0.02}$&$0.85_{\pm0.04}$&$0.81_{\pm0.04}$& 
N.A.               &$\mathbf{0.89}_{\pm0.02}$ \tstrut\\ 
\rowcolor{Gray}
RVBP       &$\mathbf{0.97}_{\pm0.02}$&$0.88_{\pm0.03}$&$0.91_{\pm0.04}$& N.A.               &$\mathbf{0.91}_{\pm0.02}$ \tstrut\\ 
LABP       &$\mathbf{0.96}_{\pm0.02}$&$0.92_{\pm0.03}$& N.A.               &$0.93_{\pm0.03}$&$\mathbf{0.94}_{\pm0.02}$\tstrut\\
\rowcolor{Gray}
RABP       &$\mathbf{0.98}_{\pm0.03}$&$0.93_{\pm0.05}$& N.A.              &$0.93_{\pm0.06}$&$\mathbf{0.94}_{\pm0.04}$\tstrut\\
\bottomrule
\end{tabular}
\end{table}

\paragraph{\textbf{Representation learning robustness.}} To simulate scenarios with incomplete or defective CMR data, we randomly remove 1 or 2 CMR planes from each subject and generate a new latent representation. Calculating the cosine similarity between this representation and the one with all planes available, we observe an average value of 1.0 in both cases, indicating a total alignment. Further, we conduct downstream LVM and RVEF predictions with 2 slices masked out and compare them with the regular setting (no maskout). The predicted value difference is $\pm 0.3$g for LVM and $\pm 0.4\%$ for RVEF, emphasizing its robustness and significant potential in practical scenarios with missing planes.

\paragraph{\textbf{Limitation and outlook.}} 
While the presented evaluation is limited to prediction and segmentation tasks, future work will further explore tasks requiring whole 3D+T information, such as cardiac motion tracking and other modality mapping (e.g. genetics) to widen the scope of cardiac analysis. While our model demonstrates meaningful representations across diverse subjects, it has not been evaluated on longitudinal data. This presents an avenue for tracking potential cardiac disease progression by observing patient trajectories in the latent space. Additionally, the capability of our method to offer comprehensive cardiac representations with fewer CMR images opens opportunities to reduce CMR scan times and alleviate patient discomfort.


\section{Conclusion}
In this study, we introduced a self-supervised method for 3D+T cardiac representation learning, utilizing multi-view sparse 2D CMR images. Trained on 14,000 CMR datasets, our approach is adaptable to various downstream tasks and maintains robust performance even with missing CMR slices. Leveraging the heart's spatiotemporal information, our model enables accurate cardiac phenotype prediction and efficient, precise whole-heart segmentation.

\section{Acknowledgements}
This research has been conducted using the UK Biobank Resource under Application Number 87802. This work is funded by the European Research Council (ERC) project Deep4MI (884622). 

\bibliographystyle{splncs04}

\bibliography{ref_short.bib}

\clearpage
\appendix
\section{Supplementary}

\begin{table}[ht!]
\centering
\setlength{\tabcolsep}{3mm}{}
\caption{Reconstruction PSNR (dB) for SA and LA views. From left to right it shows the PSNR of our model trained with only SA 2D+T planes, only LA 2D+T planes, and all available multi-view planes separately.}
\label{tab:recon}
\begin{tabular}{l|c|c|c}
\toprule
\diagbox[width=6em]{Eval.}{Input}  & SA & LA  &  ALL    \\ \hline
\rowcolor{Gray}
SA         & $26.20_{\pm0.84}$   & N.A.                & $\mathbf{31.08_{\pm0.86}}$\tstrut\\ 
LA        & N.A.                & $25.11_{\pm1.48}$    & $\mathbf{28.60_{\pm1.50}}$\tstrut\\
\bottomrule
\end{tabular}
\end{table}

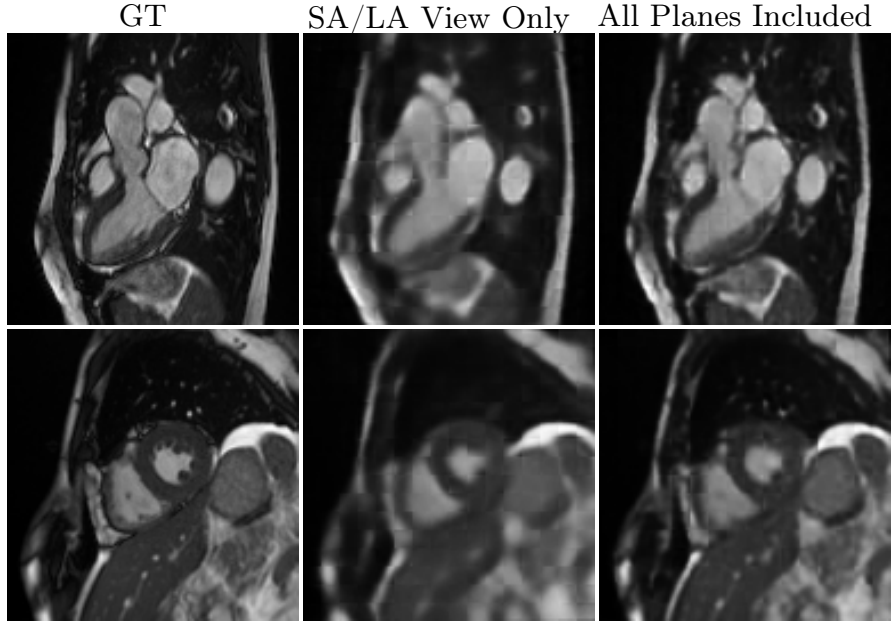
\begin{figure}[ht!]
  \centering
  \resizebox{\textwidth}{!}{%
    \begin{tikzpicture}[font=\small]
    \node[anchor=north] at (1.5, 13.5) {GT};
    \node[anchor=north] at (4.5, 13.5) {SA/LA View Only};
    \node[anchor=north] at (7.5, 13.5) {All Planes Included};
    \node[anchor=south west, scale=1.0] at (0,10)
    {\includegraphics[width=2.95cm]{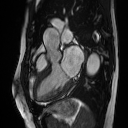}};
    \node[anchor=south west, scale=1.0] at (3,10)
    {\includegraphics[width=2.95cm]{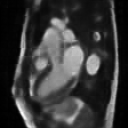}};
    \node[anchor=south west, scale=1.0] at (6,10)
    {\includegraphics[width=2.95cm]{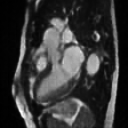}};
    \node[anchor=south west, scale=1.0] at (0,7)
    {\includegraphics[width=2.95cm]{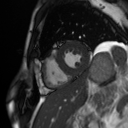}};
    \node[anchor=south west, scale=1.0] at (3,7)
    {\includegraphics[width=2.95cm]{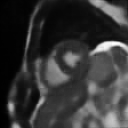}};
    \node[anchor=south west, scale=1.0] at (6,7)
    {\includegraphics[width=2.95cm]{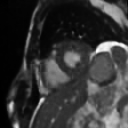}};
    \end{tikzpicture}
    }
  \caption{Reconstruction samples in the pre-training phase. The second column shows reconstructions with input planes from a single view (SA/LA only), while the last column shows the reconstructions with planes from both views included.}
  \label{fig:recon}
\end{figure}

\end{document}